\documentclass[aps,prd,twocolumn,superscriptaddress,altaffilletter,nobibnotes,nofootinbib,10pt,showpacs]{revtex4-1}
\usepackage{amsmath, amsthm, amssymb} 
\usepackage{graphicx}
\usepackage{epsfig}
\usepackage{pslatex}   
\usepackage{color}
\usepackage{endnotes}
\usepackage{hyperref}

\newcommand {\pythia} {{\sc pythia}}
\newcommand {\phojet} {{\sc phojet}}

\newcommand {\pp} {\mbox{$pp$}~}
\newcommand {\pbarp} {\mbox{$p\bar{p}$}~}
\newcommand {\ee} {\mbox{$e^+e^-$}~}

\def\dedx      {d$E$/d$x$}
\def\qinv      {$q_{\rm inv}$}
\def\rinv      {$R_{\rm inv}$}
\def\pt        {$p_{\rm T}$}
\def\bpt       {$\mathbf{p_{\rm T}}$}
\def\kt        {$k_{\rm T}$}
\def\meankt    {$\langle$\kt$\rangle$}
\def\pim       {$\pi^-$}
\def\pip       {$\pi^+$}
\def\ene       {$\sqrt{s}=900$~GeV}
\def\dndeta    {d$N_{\rm ch}$/d$\eta$}
\def\meandndeta{$\langle$\dndeta$\rangle$}
\def\deg       {$^{\rm o}$}
\def\gevc      {~GeV/$c$}

\usepackage{preprintcover}
\PreprintIdNumber{}
\PreprintCoverPaperTitle{Two-pion Bose-Einstein correlations in \pp collisions at \ene}
\PreprintCoverAbstract{  We report on the measurement of two-pion correlation functions 
  from \pp collisions at \ene\ performed by the ALICE experiment  
  at the Large Hadron Collider. 
  Our analysis shows an increase of the HBT radius with increasing 
  event multiplicity, in line with other measurements 
  done in particle- and nuclear collisions. 
  Conversely, the strong decrease of the radius with increasing  
  transverse momentum, as observed at RHIC and at Tevatron, 
  is not manifest in our data. 
}
\PreprintJournalName{PRD}

\begin{document}

\title{Two-pion Bose-Einstein correlations in \pp collisions at \ene}
\collaboration{ALICE Collaboration}
\author{K.~Aamodt}
\affiliation{Department of Physics, University of Oslo, Oslo, Norway} 
\author{N.~Abel}
\affiliation{Kirchhoff-Institut f\"{u}r Physik, Ruprecht-Karls-Universit\"{a}t Heidelberg, Heidelberg, Germany} 
\author{U.~Abeysekara}
\affiliation{Physics Department, Creighton University, Omaha, NE, United States} 
\author{A.~Abrahantes~Quintana}
\affiliation{Centro de Aplicaciones Tecnol\'{o}gicas y Desarrollo Nuclear (CEADEN), Havana, Cuba} 
\author{A.~Abramyan}
\affiliation{Yerevan Physics Institute, Yerevan, Armenia} 
\author{D.~Adamov\'{a}}
\affiliation{Nuclear Physics Institute, Academy of Sciences of the Czech Republic, \v{R}e\v{z} u Prahy, Czech Republic} 
\author{M.M.~Aggarwal}
\affiliation{Physics Department, Panjab University, Chandigarh, India} 
\author{G.~Aglieri~Rinella}
\affiliation{European Organization for Nuclear Research (CERN), Geneva, Switzerland} 
\author{A.G.~Agocs}
\affiliation{KFKI Research Institute for Particle and Nuclear Physics, Hungarian Academy of Sciences, Budapest, Hungary} 
\author{S.~Aguilar~Salazar}
\affiliation{Instituto de F\'{\i}sica, Universidad Nacional Aut\'{o}noma de M\'{e}xico, Mexico City, Mexico} 
% ...
 % from 2-Jul-2010 at 9:00

\date{\today}

\begin{abstract}
  We report on the measurement of two-pion correlation functions 
  from \pp collisions at \ene\ performed by the ALICE experiment  
  at the Large Hadron Collider. 
  Our analysis shows an increase of the HBT radius with increasing 
  event multiplicity, in line with other measurements 
  done in particle- and nuclear collisions. 
  Conversely, the strong decrease of the radius with increasing  
  transverse momentum, as observed at RHIC and at Tevatron, 
  is not manifest in our data. 
\end{abstract}

\maketitle

%%%%%%%%%%%%%%%%%%%%%%%%%%%%%%%%%%%%%%%%%%%%%%%%%%%%%%%%%%%%%%%%%%%%%%%%%%%%%%%
\section{\label{sec:intro}Introduction}
%%%%%%%%%%%%%%%%%%%%%%%%%%%%%%%%%%%%%%%%%%%%%%%%%%%%%%%%%%%%%%%%%%%%%%%%%%%%%%%

Proton-proton collisions at \ene\ have been recorded by ALICE 
(A Large Ion Collider Experiment) at the Large Hadron Collider (LHC) at 
CERN~\cite{Collaboration:2009dt}. 
Hadron collisions at these energies provide an opportunity to probe 
Quantum Chromodynamics (QCD) under extreme conditions. The
distinguishing feature of QCD is the mechanism of color confinement,
the physics of which is not fully understood, due in part to its
theoretical intractability~\cite{Peshier:2002ww}.  The confinement
mechanism has a physical scale on the order of the proton radius and
is especially important at low momentum.  

Bose-Einstein enhancement of identical-pion pairs at low relative momentum 
was first observed in \pbarp collisions by Goldhaber, Goldhaber, 
Lee and Pais 50 years ago~\cite{Goldhaber:1960sf}. 
Since then, two-pion correlations have been successfully applied 
to assess the spatial scale of the emitting source in \ee\cite{Kittel:2001zw}, 
hadron-hadron and lepton-hadron~\cite{Alexander:2003ug}, and heavy 
ion~\cite{Lisa:2005dd} collisions. Especially in the latter case,
this technique, known as Hanbury Brown - Twiss (HBT) 
interferometry~\cite{HBT1,Hanbury:1954wr} and being a special case of 
femtoscopy~\cite{Lednicky:1990pu,Lednicky:2002fq}, 
has been developed into a precision tool to
probe the dynamically-generated geometry of the emitting system.
In particular, a first order phase transition between the 
color-deconfined and -confined states was precluded by the observation 
of short timescales~\cite{Lisa:2005dd}. 
At the same time, femtoscopic measurement of bulk collective flow, 
manifesting itself via dynamical dependences of femtoscopic scales (``homogeneity 
lengths''~\cite{Akkelin:1995gh,Sinyukov:1994vg}), provided hints 
that a strongly self-interacting system was created in the collision. 
This was further corroborated by the positive correlation between the 
HBT radius and the multiplicity of the event~\cite{Lisa:2005dd}. 

In particle physics, 
overviews of femtoscopic measurements in hadron- and lepton-induced 
collisions~\cite{Kittel:2001zw,Alexander:2003ug,Chajecki:2009zg} 
reveal systematics surprisingly similar to those mentioned above
for heavy-ion collisions.  Moreover, in the first direct
comparison of femtoscopy in heavy-ion collisions at RHIC, and proton
collisions in the same apparatus, a virtually identical multiplicity-
and momentum-dependence was reported in the two 
systems~\cite{aggarwal:2010bw}. 

A systematic program of femtoscopic measurements in \pp and heavy-ion 
collisions at the LHC will shed considerable light on the nature, 
the similarities, and the differences of their dynamics. With the 
present work, we begin this program.

%%%%%%%%%%%%%%%%%%%%%%%%%%%%%%%%%%%%%%%%%%%%%%%%%%%%%%%%%%%%%%%%%%%%%%%%%%%%%%%
\section{\label{sec:exp}Experiment and data analysis}
%%%%%%%%%%%%%%%%%%%%%%%%%%%%%%%%%%%%%%%%%%%%%%%%%%%%%%%%%%%%%%%%%%%%%%%%%%%%%%%

The data discussed in this article were collected in December 2009, 
during the first stable-beam period of the LHC commissioning. 
The two beams were at the LHC injection energy of 450 GeV and each had 
2-4 bunches, one of them colliding at the ALICE intersection point. 
The bunch intensity was typically $5 \times 
10^9$ protons, giving a luminosity of the order of 
$10^{26}$~cm$^{-2}$~s$^{-1}$ and a rate for inelastic proton-proton 
collisions of a few Hz.

%\subsection{Run conditions, trigger, etc.}

Approximately $3 \times 10^5$ minimum bias \pp collision events were 
identified by signals measured in the forward scintillators (VZERO) and 
the two layers of the Silicon Pixel Detector (SPD)~\cite{Aamodt:2008zz}.
The VZERO counters are placed 
on either side of the interaction region at z = 3.3 m and z = -0.9 m. They 
cover the region $2.8<\eta<5.1$ and $-3.7<\eta<-1.7$ and record both 
amplitude and time of signals produced by charged particles. 
The minimum-bias trigger required a hit in one of the VZERO counters 
or in one of the two SPD layers which cover the central pseudorapidity 
regions $|\eta|<2$ (inner) and $|\eta|<1.4$ (outer). 
The events were collected in coincidence with the signals from two beam 
pick-up counters, one on each side of the interaction region, indicating the 
presence of passing bunches. 
The trigger selection efficiency for inelastic collisions was 
estimated to be 95-97\% \cite{Aamodt:2010ft}. 

The VZERO counters were used also to discriminate against beam-gas and 
beam-halo events by requiring a strict matching between their timing 
signals (see Ref.~\cite{Collaboration:2009dt} for details). 
This background was also rejected by 
exploiting the correlation between the number of clusters of pixels and 
the number of tracklets pointing to a reconstructed vertex. After these 
selections the fraction of background events remaining in the sample 
of events with at least one charged particle track was estimated 
to be below 0.1\%. The trigger and run conditions are discussed in detail 
in Ref.~\cite{Aamodt:2010ft}. 
  
%\subsection{Event selection}

The 250 k events used in the analysis were required to have a primary 
vertex (collision position) within 10~cm of the center of the 5~m long 
Time Projection Chamber (TPC)~\cite{Alme:2010ke}. 
This provides almost uniform acceptance for particles within the 
pseudorapidity range $|\eta|<0.8$ for all events in the sample. 
Within this sample, we have selected events based on the measured
charged-particle multiplicity $M$. 
The three multiplicity classes were $M \leq 6$, $7\leq M \leq 11$, 
and $M \geq 12$; about 70\% of all events were falling into the first 
multiplicity class. 
The tracks used in determining the multiplicity were the same as 
those used for correlation analysis (see below) except that 
particle identification cuts were not applied. 
The measured multiplicity was converted to the charged-particle 
pseudorapidity density \dndeta\ by normalizing it to the pseudorapidity 
acceptance and by correcting it for the reconstruction efficiency 
and contamination. The correction factor was determined from a Monte Carlo 
simulation with the \phojet\ event generator \cite{Engel:1994vs,Engel:1995yda} 
and with 
the full description of the ALICE apparatus and is 0.71, 0.78, and 0.81,  
respectively, for the three multiplicity bins. 
The estimated systematic error is below 4\%. 
The average charged-particle pseudorapidity density of the analyzed 
event sample is \meandndeta=3.6. 
An alternative method based on SPD tracklets \cite{Aamodt:2010ft} gave 
the same result. 
%High multiplicity events contribute more pairs; weighted with the 
%number of pairs the mean multiplicity becomes 
%$\langle($\dndeta$)^3\rangle/\langle($\dndeta$)^2\rangle$ = 9.4.

%\subsection{Track selection}

The ALICE Time Projection Chamber (TPC)~\cite{Alme:2010ke} was used to
record charged particle tracks as they leave ionization trails in the
Ne-CO$_2$-N$_2$ gas.  The ionization electrons drift up to 2.5~m to be
measured on 159~pad rows; the position resolution is better than 2~mm. 
Combined with a solenoidal magnetic field of B=0.5~T this leads to
a momentum resolution $\sim 1\%$ for pions with \pt~$<1$\gevc. 
The ALICE Inner Tracking System (ITS) has also been used for tracking. It
consists of six silicon layers, two innermost pixel detectors, two
layers of drift detectors, and two outer layers of strip detectors,
which provide up to six space points for each track. The tracks used
in this analysis were reconstructed using the information from both 
the TPC (signals from at least 90 pad rows required) and the ITS. 
Separate studies have been done with TPC-only and ITS-only tracks, 
and were found to give results consistent with the combined ITS+TPC analysis.
The tracks were required to project back to the primary interaction
vertex within $0.2$~cm (2.4~cm) in the transverse plane and $0.25$~cm 
(3.2~cm) in longitudinal direction, if ITS+TPC (TPC only) information 
is used, thereby rejecting most secondary pions from weak decays. 
The pion tracks used in the correlation analysis had transverse 
momenta between 0.15\gevc\ and 1.0~\gevc. 

ALICE provides excellent particle identification capability. 
In this analysis the particle identification was achieved by correlating 
the magnetic rigidity of a track with its specific ionization (\dedx) 
in the TPC gas. The \dedx\ of the TPC was calibrated using cosmic rays 
and its resolution was shown to be better than 5.5\%, the design value. 
The contamination of the pion sample is negligible within the momentum
range of 0.25\gevc~$<p<$~0.65\gevc. Below and above this range it is 
on the order of 5\% and is caused by electrons and kaons, respectively. 

%%%%%%%%%%%%%%%%%%%%%%%%%%%%%%%%%%%%%%%%%%%%%%%%%%%%%%%%%%%%%%%%%%%%%%%%%%%%%%%
\section{\label{sec:cor}Two-pion correlation functions}
%%%%%%%%%%%%%%%%%%%%%%%%%%%%%%%%%%%%%%%%%%%%%%%%%%%%%%%%%%%%%%%%%%%%%%%%%%%%%%%

%\subsection{Pairing, pair selection, correlation functions}

The two-particle correlation function is defined as the ratio 
$C\left({\bf q}\right)=A\left({\bf q}\right)/B\left({\bf q}\right)$, 
where  $A\left({\bf q}\right)$ 
is the measured distribution of pair momentum difference 
${\bf q}={\bf p}_2-{\bf p}_1$, and $B\left({\bf q}\right)$ is a similar
distribution formed by using pairs of particles from different
events (event mixing)~\cite{Kopylov:1974th}. 
The limited statistics available (520 k identical-pion pairs with 
\qinv~$<$~0.5\gevc) allowed us to perform
a detailed analysis only for the one-dimensional two-pion correlation 
function $C$(\qinv). The \qinv\ is, for identical mass particles, equal to 
the modulus of the momentum difference $|{\bf q}|$ in the pair rest frame. 
% --(0-3)  686697 --(0-0.5) 265154
% ++(0-3)  677240 ++(0-0.5) 257865
% -+(0-3) 1488176 -+(0-0.5) 564420

The correlation functions were studied in bins of event multiplicity 
and of transverse momentum, defined as half of the vector 
sum of the two transverse momenta, \kt~=~$|$\bpt$_{,1}$+\bpt$_{,2}|/2$. 
During event mixing, all pion tracks from one event were paired 
with all pion tracks from another event. 
Every event was mixed with five other events with similar multiplicities; 
ten multiplicity bins were introduced for this purpose. 
The multiplicity binning improved the flatness of the correlation function 
at \qinv~$>$~1.5\gevc. 
Binning events according to their vertex position, on the other hand, 
had no effect on the correlation function and therefore was not used. 
Alternatively to event mixing, the denominator can be obtained by 
rotating one of the two tracks by 180\deg\ in azimuth. 
The correlation functions obtained using this technique are generally 
flatter at high \qinv\ than those from event mixing. The difference 
between the results obtained utilizing the two techniques was 
used in estimating the systematic errors. 

For the correlation structures measured here, with characteristic
widths $\sim 0.2$\gevc, track splitting and track merging 
in the event reconstruction are small effects overall.  
Their impact on the results was carefully studied with the Monte 
Carlo simulation and turned out to be negligible. 

Another apparatus effect considered is the momentum
resolution. Momentum smearing for single particles has similar effect
on the correlation structures in two-particle correlations i.e. it smears
the correlation peak, making it appear lower and wider. 
We have studied this effect with the Monte-Carlo simulation of
the ALICE detector and have found that for the width of the 
correlation peak expected here the effect is on the order of 1\%.  

%\subsection{Identical-pion correlation functions}

\begin{figure}[b]
\hspace*{-2mm}\includegraphics{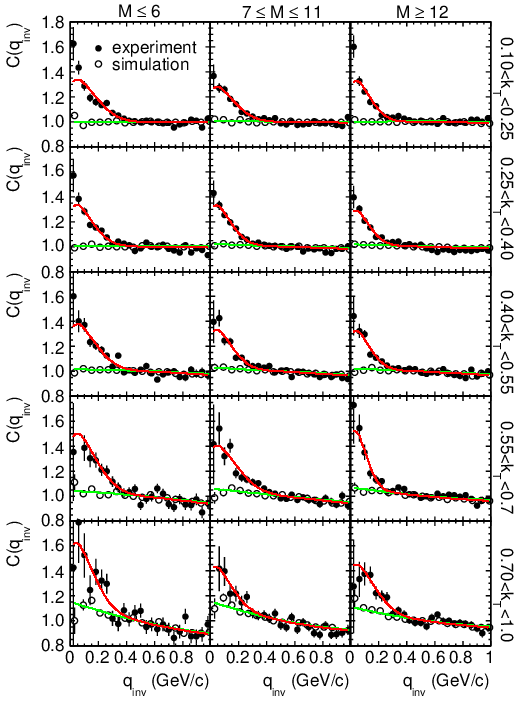}
\caption{
Correlation functions for identical pions from \pp collisions 
at \ene\ (full dots) and those obtained from a simulation 
using \phojet\ (open circles). 
Positive and negative pion pairs were combined. 
The three columns represent collisions with different charged-particle 
multiplicities $M$; the transverse momentum of pion pairs 
\kt~(GeV/c) increases from top to bottom. 
The lines going through the points represent the Gaussian fits discussed in 
the text. 
\label{fig:ident}}
\end{figure}

Fig.~\ref{fig:ident} presents two-pion correlation functions measured by 
ALICE in \pp collisions at \ene, as a function of event multiplicity and 
transverse momentum \kt. 
The denominator of the correlation function was obtained via 
event mixing and normalized such that the numbers of true and mixed pairs 
with 0.4\gevc~$<$~\qinv~$<$~0.6\gevc\ were equal. 
The \qinv\ range used for normalization was chosen to be outside 
of the Bose-Einstein peak but as close as possible to it. 
The normalized distributions of positive and negative pion pairs were 
added together before building the ratio of true and mixed pairs. 
The Bose-Einstein enhancement is manifest at low \qinv. 
A slight decrease of the correlation peak width is seen as multiplicity 
grows. The \kt\ dependence is less obvious because the correlation 
baseline -- the underlying two particle correlation without any 
Bose-Einstein enhancement -- is systematically changing its shape 
between the low and high transverse momenta. 

The correlation functions were fitted by a function accounting for the 
Bose-Einstein enhancement and for the mutual Coulomb interaction 
between the two particles: 
\begin{equation}
\label{eq:1dbowler-sinyukov}
C(q_{inv})=\left( (1-\lambda)+
\lambda K(q_{inv}) \left[1 + {\rm
  exp}(-R_{inv}^2q_{inv}^2)\right]\right) \ D(q_{inv}) ,
\end{equation}
with $\lambda$ describing the correlation strength and \rinv\ being 
the Gaussian HBT radius \cite{Sinyukov:1998fc}.
The factor $K$ is the Coulomb function integrated over a spherical source
of the size 1~fm. It is attenuated by the same factor $\lambda$ as the 
Bose-Einstein peak. 
The factor $D$(\qinv) accounts for long-range correlations, like those  
arising from jets  and/or from energy and momentum conservation, 
and plays an important role in the analysis as will be discussed 
later. 

Neglecting the Coulomb interaction $K(q_{inv})\equiv 1$ the fit 
function reduces to 
\begin{equation}
\label{eq:gausfit}
C(q_{inv})=\left[1+\lambda \ \exp(-R_{inv}^2q_{inv}^2)\right]
\ D(q_{inv}) \ . 
\end{equation}
The difference between the \rinv\ values obtained 
with and without the Coulomb correction is less than 0.05~fm. 

While the Gaussian fit captures the bulk scales of the correlation, 
at low \qinv\ the data points lie above the fit line. 
This feature was observed previously in pion correlations from particle 
collisions. An exponential fit 
\begin{equation}
\label{eq:expfit}
C(q_{inv})=\left[1+\lambda \ \exp(-R_{inv}q_{inv})\right]
\ D(q_{inv}) 
\end{equation}
matches the data better. However, contrary to the Gaussian \rinv, 
the \rinv\ parameter from Eq.~(\ref{eq:expfit}) does not have a 
straightforward interpretation as the ``radius of the source''. 
We have used both functional forms and leave a detailed
investigation of the correlation peak shape to future studies.
In order to make the connection to established systematics at lower 
energy particle and heavy-ion collisions, a careful treatment of the 
long-range correlations, visible as a slope in the baseline of the 
correlation developing with increasing transverse momentum and 
represented by the factor $D$(\qinv) in Eqs. 
%(\ref{eq:1dbowler-sinyukov}) and (\ref{eq:gausfit}), is crucial. 
(\ref{eq:1dbowler-sinyukov}-\ref{eq:expfit}), is crucial. 

In order to better understand the shape of the correlation baseline 
we have calculated correlation functions for \pp collisions events 
generated by the model \phojet\ and propagated through the ALICE 
detectors, performing an identical analysis for 
the simulated events as for the measured ones. 
The results are shown as open circles in 
Fig.~\ref{fig:ident}. 
The model does not contain the Bose-Einstein effect, hence the lack 
of the peak at low \qinv\ is expected. At low \kt\ and low multiplicity, 
the model predicts a flat correlation function. 
However, as \kt\ increases, long-range correlations start becoming 
visible as a distortion of the correlation function baseline similar to 
that seen in the experimental data.  

%\subsection{Nonidentical-pion correlation functions}

\begin{figure}[t]
\hspace*{-2mm}\includegraphics{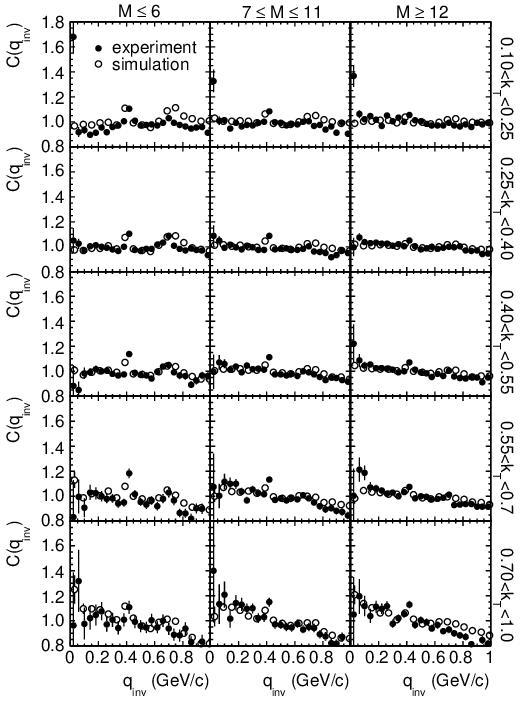}
\caption{
One-dimensional correlation functions for \pip\pim\ pairs 
from \pp collisions at \ene. 
The columns and rows are defined as in Fig.~\ref{fig:ident}.
\label{fig:nonident}}
\end{figure}

The accuracy of our simulation in describing the correlation baseline was 
verified with unlike-sign pion pairs. 
The multiplicity and \kt\ dependence of the $\pi^+ \pi^-$ functions is 
shown in Fig.~\ref{fig:nonident}. 
Correlation structures for non-identical pions include a mutual
Coulomb interaction peak, here limited to the first bin at lowest
\qinv, and peaks coming from meson decays which should be
correctly modeled in the event generator. 
Therefore, one can directly compare simulations with data. 
In Fig.~\ref{fig:nonident}, the simulated correlation functions agree 
reasonably well with the experimental data. This suggests that the 
same model (\phojet) can be used as a reasonable estimate also for 
identical particles to describe the correlation baseline under the 
Bose-Einstein peak. 
The presence of resonance peaks (like the $K^{0}_{S}$ one at 
\qinv~=~412 MeV/$c$) and the fact that the simulated correlations 
for identical and non-identical pion pairs have different slopes, on the 
other hand, indicate that unlike-sign pion pairs cannot be directly used for 
the denominator of the identical pion correlations. 

%\subsection{HBT radius}

The procedure employed to extract the HBT radii with 
Eq.~(\ref{eq:1dbowler-sinyukov}) using the \phojet\ baseline is as follows. 
First, the simulation points shown in Fig.~\ref{fig:ident} are fitted 
with the 2nd-order polynomial 
\begin{equation}
D(q_{inv}) = a + b q_{inv} + c q_{inv}^2 \ .
\label{eq:BMC}
\end{equation}
Subsequently, the experimental correlation function is fitted by 
Eq.~(\ref{eq:1dbowler-sinyukov}), taking the $D$(\qinv) from the 
\phojet\ fit and adjusting $\lambda$ and \rinv. 
The two fits are represented in Fig.~\ref{fig:ident} by the lines going 
through the simulation and experiment data points, respectively. 

In order to estimate the systematic error from the baseline determination 
we repeated the fitting procedure using a simulation performed with 
the \pythia\ \cite{Sjostrand:2006za} generator (version 6.4.21, Perugia-0 
(320) tune~\cite{Skands:2009zm}) instead of \phojet. 
The HBT radii obtained in the two ways differ by up to 10\%. In the 
following we use the average between them and we estimate the systematic 
error related to the baseline shape assumption to be half of the difference. 

It is interesting to see what happens with the radii if the slope of 
the baseline is neglected. Assuming a flat baseline $D($\qinv$)\equiv a$ and 
treating $a$ as the third fit parameter in Eq.~(\ref{eq:1dbowler-sinyukov}) 
leads to \rinv\ values that are similar to those obtained with the 
\phojet\ or \pythia\ 
baseline at low \kt\ values but smaller by up to 30\% at high transverse 
momenta. This is because the broad enhancement caused by long-range 
correlations will be attributed to Bose-Einstein correlations, giving rise
to smaller radii (wider correlation function). The resulting apparent 
\kt\ dependence will be discussed in Section~\ref{sec:ktdep}. 

\begin{figure}[h]
\centerline{\includegraphics[width=0.45\textwidth]{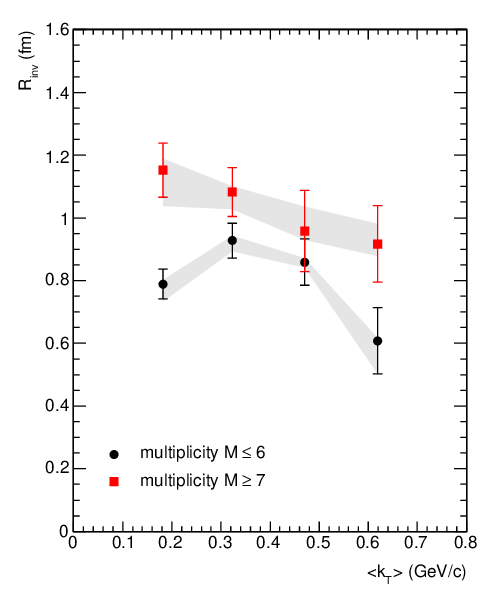}}
\caption{\label{fig:radmkdep}
Extracted HBT radius as a function of \kt\ for low (black circles) 
and high (red squares) multiplicity events. The error bars are statistical. 
The shaded bands represent the systematic errors related 
to the baseline shape assumption and to the fit range, added quadratically. 
}
\end{figure}

The \rinv\ obtained from the fit (the two highest multiplicity bins 
combined) is shown in Fig.~\ref{fig:radmkdep}. 
In order to reduce the statistical errors and to compare 
to other experiments, in the following sections of this article we analyze 
separately the multiplicity and the transverse momentum dependences. 

%%%%%%%%%%%%%%%%%%%%%%%%%%%%%%%%%%%%%%%%%%%%%%%%%%%%%%%%%%%%%%%%%%%%%%%%%%%%%%%
\section{\label{sec:mudep}Multiplicity dependence of the HBT radius}
%%%%%%%%%%%%%%%%%%%%%%%%%%%%%%%%%%%%%%%%%%%%%%%%%%%%%%%%%%%%%%%%%%%%%%%%%%%%%%%

\begin{figure}[h]
\centerline{\includegraphics{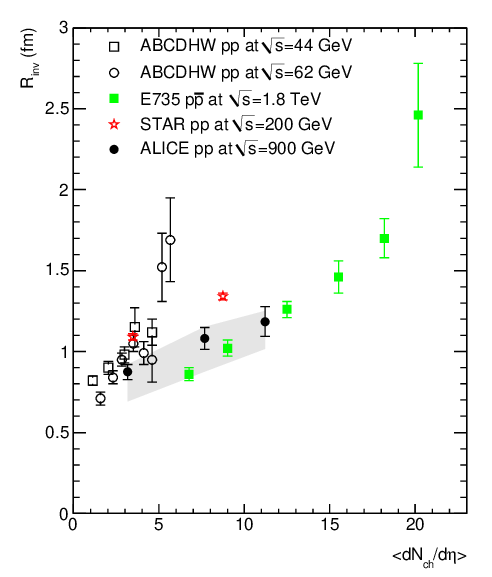}}
\caption{
One-dimensional Gaussian HBT radius in \pp collisions at \ene\, 
determined using pion pairs with \kt~=~0.1-0.55\gevc, \meankt = 0.32\gevc, 
and shown as a function of the charged-particle multiplicity at midrapidity 
(full dots). 
The shaded band represents the systematic errors (see text). 
For comparison, open symbols, red stars, and green filled boxes represent 
the data taken at the ISR~\cite{Breakstone:1986xs}, 
RHIC~\cite{aggarwal:2010bw}, 
and Tevatron~\cite{Alexopoulos:1992iv}, respectively. 
\label{fig:mudep}}
\end{figure}

The multiplicity dependence of the obtained HBT radius is shown 
in Fig.~\ref{fig:mudep} and Table~\ref{tab:mudep}. 
\begin{table}[h]
\caption{\label{tab:mudep}
One-dimensional HBT radius in \pp collisions at \ene\, 
determined using pion pairs with \kt~=~0.1-0.55\gevc, 
\meankt = 0.32\gevc, 
as a function of the charged-particle pseudorapidity density at midrapidity. 
The radii were obtained using the Gaussian fit function 
defined by Eq.~(\ref{eq:1dbowler-sinyukov}).}
\begin{ruledtabular}
\begin{tabular}{ccc}
\meandndeta &      $\lambda$      &  \rinv (fm) \\
\colrule
   3.2  &  0.386 $\pm$ 0.022  &  0.874 $\pm$ 0.047 (stat.) $^{+ 0.047}_{- 0.181}$ (syst.) \\
   7.7  &  0.331 $\pm$ 0.023  &  1.082 $\pm$ 0.068 (stat.) $^{+ 0.069}_{- 0.206}$ (syst.) \\
  11.2  &  0.310 $\pm$ 0.026  &  1.184 $\pm$ 0.092 (stat.) $^{+ 0.067}_{- 0.168}$ (syst.) \\
\end{tabular}
\end{ruledtabular}
\end{table}
The analysis here was restricted to the first three transverse momentum 
bins \kt~=~0.1-0.55\gevc. The mean transverse momentum for pairs with 
\qinv~$<$~0.2\gevc\ is \meankt = 0.32\gevc. 
The HBT radii were obtained by using \phojet\ and \pythia\ to estimate 
the shape of the baseline, as explained in the previous section. 
The systematic errors related to the baseline assumption reflect the 
difference between the two. 
The systematic error related to the choice of the normalization and/or fit 
range was estimated to be 5\%. 
An additional downward systematic error of 13-20\% accounts for the difference 
between the event mixing and the rotation denominator techniques. 
The shaded area represents the three systematic errors added in 
quadrature. 

The charged-particle pseudorapidity density \meandndeta\ 
of the lowest multiplicity bin was calculated excluding events with 
multiplicities $M<2$ because these events do not contribute to the 
numerator of the correlation function. Including all events and including 
only events with at least one like-sign pair would shift the point by 0.8 
to the left and to the right, respectively. 

An increase of the HBT radius with multiplicity is observed, 
consistent with the hadron-hadron collision systematics above 
$\sqrt{s}\sim 50$~GeV~\cite{Chajecki:2009zg}. 
While the average transverse momentum is similar in all four data 
sets, other aspects of the analysis, e.g. the average orientation of the 
momentum difference vector, can differ so the trends, not the
absolute values, should be compared. In heavy-ion collisions, this
multiplicity dependence has been associated with the particle composition 
and overall volume of the final state 
system~\cite{Lisa:2005dd,Antonczyk:2007id,Adamova:2002ff} 
or with final-state hadronic rescattering~\cite{Humanic:2008nt}. 
The relation observed in heavy-ion collision data~\cite{Lisa:2005dd},
$R \sim a + b($\dndeta$)^{1/3}$, 
where $a$ and $b$ are constants, appears to be consistent with our 
data within our systematic errors. 
For high energy \pp collisions, it has been suggested that a similar 
behavior could originate from final-state hadronic rescattering for 
short hadronization times~\cite{Humanic:2006ib}. In an alternative 
scenario, the increase of the HBT radius with multiplicity results 
from the fact that the high multiplicity \pp events mostly come from 
hard parton scattering, and the hadronization length, i.e. the distance 
travelled by a parton before hadronization, is roughly proportional 
to the parton energy~\cite{Paic:2005cx}. 

The fitted correlation strength $\lambda$ is lower than unity, the value 
expected for the ideal Bose-Einstein case. 
One reason for this is the non-Gaussian shape of the peak, caused at 
least partially by pions from decays of short- ($\Delta$, $\rho$) and 
long-lived resonances ($\omega$, $\eta$, $\eta'$). On the detector 
side, $\lambda$ can be reduced by the particle misidentification; this 
effect is however small in our data sample. 
In ALICE, $\lambda$ decreases from 0.37$\pm$0.03 to 0.32$\pm$0.03 between 
the lowest and the highest multiplicity, in close agreement with the E735 
measurements at Tevatron~\cite{Alexopoulos:1992iv}. 
A similar trend was observed by UA1 in \pbarp collisions at 
$\sqrt{s} = 630$\gevc\ \cite{Albajar:1989sj}; 
the fact that their $\lambda$ values were lower may have to do with the 
lack of the particle identification and the resulting dilution of the 
correlation peak. 
In a final-state hadronic rescattering model~\cite{Humanic:2006ib}, 
a correlation strength dropping with multiplicity 
in high-energy \pp collisions was attributed to the increased 
contribution from long-lived resonances in higher multiplicity events. 

An increase of the HBT radius with increasing particle multiplicity 
was recently reported by the CMS Collaboration for the same collision system 
and energy \cite{Khachatryan:2010un}. 
The authors fit the correlation peak by an exponential (Eq.~(\ref{eq:expfit})). 
An analogous approach in our case (Table~\ref{tab:mudepexp}) yields radii 
that are rather similar to the Gaussian ones (Table~\ref{tab:mudep})  
once scaled down by $\sqrt{\pi}$ \cite{Khachatryan:2010un}. 
In order to compare between the two experiments we perform a fit to an 
inclusive correlation (all multiplicities and \kt's). 
The exponential fit to the correlation functions obtained using event 
mixing and using rotation yields \rinv=1.61$\pm$0.07 (stat.)$\pm$0.05 (syst.)~fm 
and \rinv=1.31$\pm$0.05 (stat.)$\pm$0.22 (syst.)~fm, respectively. This is in close 
agreement with the corresponding values quoted by CMS, 1.72$\pm$0.06~fm and 
1.29$\pm$0.04~fm. 
\begin{table}[h]
\caption{\label{tab:mudepexp}
One-dimensional HBT radius in \pp collisions at \ene\, 
determined using pion pairs with \kt~=~0.1-0.55\gevc, 
\meankt = 0.32\gevc, 
as a function of the charged-particle pseudorapidity density at midrapidity. 
The radii were obtained using the exponential fit function 
defined by Eq.~(\ref{eq:expfit}).}
\begin{ruledtabular}
\begin{tabular}{ccc}
\meandndeta &      $\lambda$      &  \rinv/$\sqrt{\pi}$ (fm)\\
\colrule
   3.2  &  0.704 $\pm$ 0.048  &  0.809 $\pm$ 0.061 (stat.) $^{+ 0.049}_{- 0.208}$ (syst.) \\
   7.7  &  0.577 $\pm$ 0.054  &  0.967 $\pm$ 0.095 (stat.) $^{+ 0.071}_{- 0.206}$ (syst.) \\
  11.2  &  0.548 $\pm$ 0.051  &  1.069 $\pm$ 0.104 (stat.) $^{+ 0.063}_{- 0.203}$ (syst.) \\
\end{tabular}
\end{ruledtabular}
\end{table}

%%%%%%%%%%%%%%%%%%%%%%%%%%%%%%%%%%%%%%%%%%%%%%%%%%%%%%%%%%%%%%%%%%%%%%%%%%%%%%%
\section{\label{sec:ktdep}Transverse momentum dependence of the HBT
radius}
%%%%%%%%%%%%%%%%%%%%%%%%%%%%%%%%%%%%%%%%%%%%%%%%%%%%%%%%%%%%%%%%%%%%%%%%%%%%%%%

One of the key features of the bulk system created in nuclear
collisions is its large collective flow. The fingerprint
of this flow is a specific space-momentum correlation signature,
revealed in the transverse momentum dependence of the Gaussian 
HBT radius~\cite{Lisa:2005dd}.  While quantitative comparison between
particle and heavy ion studies is complicated by experiments using
different acceptances and techniques, a recent comparison of 
the HBT radii from \pp and Au+Au collisions at RHIC indicates 
an almost identical \pt\ dependence between these collision 
systems~\cite{aggarwal:2010bw}. 
Again, this raises the interesting question whether hadron collisions 
at the highest energies already develop a bulk, collective behavior. 

\begin{figure}[!h]
\includegraphics{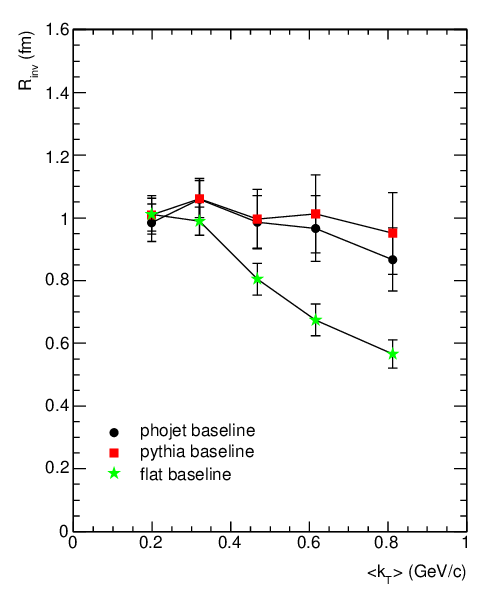}
\vspace{-5mm}
\caption{\label{fig:ktdep3}
One-dimensional Gaussian HBT radius in \pp collisions at 
\ene\ as a function of transverse momentum \kt. 
Three fitting methods, differing by the choice of the baseline 
parametrization, are compared.}
\end{figure}

The \kt\ dependence of our measured HBT radius is shown in 
Fig.~\ref{fig:ktdep3}. 
The choice of the fitting method, which only weakly affects the 
multiplicity dependence of the HBT radius discussed in the previous section, 
is of crucial importance for the transverse momentum dependence. 
Taking the baseline shape from the Monte Carlo leads to an HBT 
radius that is nearly independent of \kt\ (filled black circles 
and red boxes for \phojet\ and \pythia, respectively). 
Assuming a flat baseline, on the other hand, results in a radius falling 
with \kt\ (green stars). 
As discussed in the previous section, the experimental unlike-sign pion 
correlation functions are close to the predictions of \phojet\ and 
\pythia\ and we consider using the average between the
two cases as baseline to be a reliable estimate for the HBT radii.

\begin{figure}[h]
\includegraphics{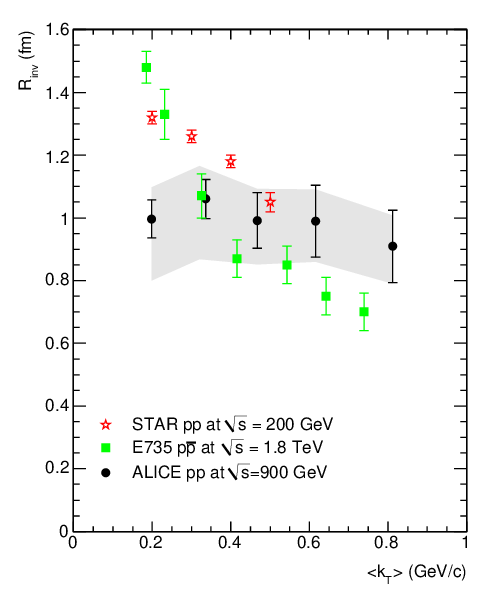}
\caption{\label{fig:ktdep}
One-dimensional Gaussian HBT radius in \pp collisions at 
\ene\ as a function of transverse momentum \kt\ (full dots). 
The mean charged-particle multiplicity density was 
\meandndeta=3.6. 
\phojet\ simulation was used to determine the baseline of the correlations. 
Using \pythia\ and a flat baseline leads to systematic deviations up and down, 
respectively; the related systematic errors as indicated by the shaded area. 
Stars and filled boxes represent the radii measured at 
RHIC~\cite{aggarwal:2010bw} and 
Tevatron~\cite{Alexopoulos:1992iv}, respectively. }
\end{figure}

The radii obtained in this fashion are summarized in Tables~\ref{tab:ktdep} 
and \ref{tab:ktdepexp} 
\begin{table}[b]
\caption{\label{tab:ktdep}
One-dimensional HBT radius in \pp collisions at \ene\ as a function of the pair \kt. 
The radii were obtained using the Gaussian fit function defined by 
Eq.~(\ref{eq:1dbowler-sinyukov}).}
\begin{ruledtabular}
\begin{tabular}{ccc}
\meankt\ (\!\gevc) &  $\lambda$  &  \rinv (fm) \\
\colrule
 0.20  &  0.35 $\pm$ 0.03  &  1.00 $\pm$ 0.06 (stat.) $^{+ 0.10}_{- 0.20}$ (syst.) \\
 0.32  &  0.33 $\pm$ 0.03  &  1.06 $\pm$ 0.06 (stat.) $^{+ 0.11}_{- 0.19}$ (syst.) \\
 0.47  &  0.30 $\pm$ 0.04  &  0.99 $\pm$ 0.09 (stat.) $^{+ 0.10}_{- 0.14}$ (syst.) \\
 0.62  &  0.35 $\pm$ 0.06  &  0.99 $\pm$ 0.11 (stat.) $^{+ 0.10}_{- 0.13}$ (syst.) \\
 0.81  &  0.31 $\pm$ 0.06  &  0.91 $\pm$ 0.12 (stat.) $^{+ 0.10}_{- 0.12}$ (syst.) \\
\end{tabular}
\end{ruledtabular}
\end{table}
\begin{table}[h]
\caption{\label{tab:ktdepexp}
One-dimensional HBT radius in \pp collisions at \ene\ as a function of the pair \kt. 
The radii were obtained using the exponential fit function defined by 
Eq.~(\ref{eq:expfit}).}
\begin{ruledtabular}
\begin{tabular}{ccc}
\meankt (\!\gevc) &  $\lambda$  &  \rinv / $\sqrt{\pi}$ (fm) \\
\colrule
 0.20  &  0.63 $\pm$ 0.05  &  0.94 $\pm$ 0.07 (stat.) $^{+ 0.09}_{- 0.20}$ (syst.) \\
 0.32  &  0.58 $\pm$ 0.04  &  0.93 $\pm$ 0.07 (stat.) $^{+ 0.09}_{- 0.20}$ (syst.) \\
 0.47  &  0.55 $\pm$ 0.07  &  0.92 $\pm$ 0.10 (stat.) $^{+ 0.09}_{- 0.14}$ (syst.) \\
 0.62  &  0.70 $\pm$ 0.11  &  0.98 $\pm$ 0.14 (stat.) $^{+ 0.10}_{- 0.14}$ (syst.) \\
 0.81  &  0.60 $\pm$ 0.12  &  0.90 $\pm$ 0.16 (stat.) $^{+ 0.12}_{- 0.15}$ (syst.) \\
\end{tabular}
\end{ruledtabular}
\end{table}
and shown in Fig.~\ref{fig:ktdep} where we compare 
them to RHIC and Tevatron data~\cite{Chajecki:2009zg}. 
Like for the multiplicity dependence, the systematic error band 
represents a quadratic sum of the error related to the baseline assumption 
(0-10\%), the fit range (10\%), and the denominator construction method 
(mixing/rotating, 7-17\%). 
The lowest-\kt\ point is significantly below the RHIC and Tevatron results. 
It should be noted that the ALICE analysis was performed on a minimum 
bias event sample and the averaged charged-particle pseudorapidity 
density is \meandndeta=3.6 while the Tevatron 
events are biased to high multiplicity, \meandndeta=14.4, 
similar to our highest multiplicity bin. 
As visible in 
Fig.~\ref{fig:radmkdep}, the lowest-\kt\ point at the high multiplicity 
is at \rinv$\approx$1.2~fm, approaching the Tevatron points. 
The STAR results, on the other hand, were obtained from events with 
\meandndeta = 4.3 i.e. similar to the 
ALICE case and thus a similar reasoning cannot explain the difference. 

Two tests were performed to make sure that the low HBT radius value at 
low transverse momenta is not caused by apparatus effects. 
First, the analysis was repeated using only the ITS and thus reducing 
the low-momentum cut-off by about 50~MeV. 
This analysis yielded the same HBT radius which demonstrates that the 
energy loss is not an issue. 
Second, 
as seen in Fig.~\ref{fig:radmkdep} the low-\kt\ point is mostly driven 
down by the contribution of the low multiplicity events. Since the vertex 
resolution in these events is worse this might in principle deteriorate 
the momentum resolution and smear out the correlation function peak. 
In order to test this the analysis was performed without using the event 
vertex constraint for momentum determination. The results, again, were 
unchanged. This, and the distinct $K^0_S$ peak in the unlike-sign pion 
correlation functions in the low-multiplicity low-\kt\ bin of 
Fig.~\ref{fig:nonident}, indicate that the momentum resolution is not 
spoiled in low multiplicity events. 

Even more important than the position of the first point, albeit related 
to it, is the question of the slope of the points in Fig.~\ref{fig:ktdep}. 
Our measured HBT radius is practically independent 
of \kt\ within the studied transverse momentum range. 
The slope crucially depends on the baseline shape assumption, as 
was shown in Fig.~\ref{fig:ktdep3}. 
The results from the experiments to which we are comparing in 
Fig.~\ref{fig:ktdep} were extracted using a flat background (although 
the STAR experiment also studied the effects of using other types of 
backgrounds for their data to account for the non-femtoscopic 
effects~\cite{aggarwal:2010bw}). 
Assuming that \phojet\ and \pythia\ are correct such a procedure may lead 
to a misinterpretation of the low-q enhancement of the correlation function, 
that is coming from long-range correlations (most probably mini-jet 
like), as a Bose-Einstein enhancement. 
As the impact of this may depend on the details of each experiment (certainly 
on the collision energy) we do not attempt to resolve this question 
quantitatively. However, we stress again the usefulness of non-identical 
pion correlation in constraining the correlation baseline. 

%%%%%%%%%%%%%%%%%%%%%%%%%%%%%%%%%%%%%%%%%%%%%%%%%%%%%%%%%%%%%%%%%%%%%%%%%%%%%%%
\section{\label{sec:summary}Summary}
%%%%%%%%%%%%%%%%%%%%%%%%%%%%%%%%%%%%%%%%%%%%%%%%%%%%%%%%%%%%%%%%%%%%%%%%%%%%%%%

In summary, ALICE has measured two-pion correlation functions in \pp
collisions at \ene\ at the LHC. Consistent with previous
measurements of high-energy hadron-hadron and nuclear collisions,
the extracted HBT radius \rinv\ increases with event multiplicity.  
Less consistent is the relation between \rinv\ and the pion transverse 
momentum where the ALICE measured HBT radius in minimum bias events is 
practically constant within our errors and within the transverse momentum 
range studied. 

\begin{acknowledgements}

%\section*{Acknowledgements}

%The ALICE collaboration would like to thank the many physicists and engineers who %participated in the preparation and
%earlier stages of this publication, in particular  S.~Bagnasco, M.~Burns, L.~Gaudichet, %C.~J{\o}rgensen, A.B.~Kaidalov, J.B.~Kinson,
%P.~Saiz, H.~Sann, and W.~Snoeys and in addition the following technicians who played an %essential role in the preparation of the detector A.~Junique,
%J.-C.~Labbe, Y.~Lesenechal and M.~Morel. We would like also to express our gratitude to %our colleagues from the
%CERN accelerator sector for preparing and running the LHC.

The ALICE collaboration would like to thank all its engineers and technicians for their invaluable contributions to the construction of the experiment and the CERN accelerator teams for the outstanding performance of the LHC complex.

The ALICE collaboration acknowledges the following funding agencies for their support in building and
running the ALICE detector:
\begin{itemize}
\item{}
Calouste Gulbenkian Foundation from Lisbon and Swiss Fonds Kidagan, Armenia;
\item{}
Conselho Nacional de Desenvolvimento Cient\'{\i}fico e Tecnol\'{o}gico (CNPq), Financiadora de Estudos e Projetos (FINEP),
Funda\c{c}\~{a}o de Amparo \`{a} Pesquisa do Estado de S\~{a}o Paulo (FAPESP);
\item{}
National Natural Science Foundation of China (NSFC), the Chinese Ministry of Education (CMOE)
and the Ministry of Science and Technology of China (MSTC);
\item{}
Ministry of Education and Youth of the Czech Republic;
\item{}
Danish Natural Science Research Council, the Carlsberg Foundation and the Danish National Research Foundation;
\item{}
The European Research Council under the European Community's Seventh Framework Programme;
\item{}
Helsinki Institute of Physics and the Academy of Finland;
\item{}
French CNRS-IN2P3, the `Region Pays de Loire', `Region Alsace', `Region Auvergne' and CEA, France;
\item{}
German BMBF and the Helmholtz Association;
\item{}
Hungarian OTKA and National Office for Research and Technology (NKTH);
\item{}
Department of Atomic Energy and Department of Science and Technology of the Government of India;
\item{}
Istituto Nazionale di Fisica Nucleare (INFN) of Italy;
\item{}
MEXT Grant-in-Aid for Specially Promoted Research, Ja\-pan;
\item{}
Joint Institute for Nuclear Research, Dubna;
\item{}
Korea Foundation for International Cooperation of Science and Technology (KICOS);
\item{}
CONACYT, DGAPA, M\'{e}xico, ALFA-EC and the HELEN Program (High-Energy physics Latin-American--European Network);
\item{}
Stichting voor Fundamenteel Onderzoek der Materie (FOM) and the Nederlandse Organisatie voor Wetenschappelijk Onderzoek (NWO), Netherlands;
\item{}
Research Council of Norway (NFR);
\item{}
Polish Ministry of Science and Higher Education;
\item{}
National Authority for Scientific Research - NASR (Autoritatea Na\c{t}ional\u{a} pentru Cercetare \c{S}tiin\c{t}ific\u{a} - ANCS);
\item{}
Federal Agency of Science of the Ministry of Education and Science of Russian Federation, International Science and
Technology Center, Russian Acedemy of Sciences, Russian Federal Agency of Atomic Energy, Russian Federal Agency for Science and Innovations and CERN-INTAS;
\item{}
Ministry of Education of Slovakia;
\item{}
CIEMAT, EELA, Ministerio de Educaci\'{o}n y Ciencia of Spain, Xunta de Galicia (Conseller\'{\i}a de Educaci\'{o}n),
CEA\-DEN, Cubaenerg\'{\i}a, Cuba, and IAEA (International Atomic Energy Agency);
\item{}
Swedish Reseach Council (VR) and Knut $\&$ Alice Wallenberg Foundation (KAW);
\item{}
Ukraine Ministry of Education and Science;
\item{}
United Kingdom Science and Technology Facilities Council (STFC);
%\item{}
%Ohio Supercomputer Center, United States;
\item{}
The United States Department of Energy, the United States National
Science Foundation, the State of Texas, and the State of Ohio.
%\item{}
%Texas Learning and Computation Center at the University of Houston, %United States.
\end{itemize}
\end{acknowledgements}

\bibliography{Citations3}
\end{document}